\documentclass[fleqn,usenatbib,onecolumn]{mnras}

\usepackage{newtxtext,newtxmath}
\usepackage[T1]{fontenc}

\DeclareRobustCommand{\VAN}[3]{#2}
\let\VANthebibliography\thebibliography
\def\thebibliography{\DeclareRobustCommand{\VAN}[3]{##3}\VANthebibliography}

\usepackage{graphicx}	% Including figure files
\usepackage{amsmath}	% Advanced maths commands

\def \bv{{\mathbfit{v}}}

\def \br{{\mathbfit{r}}}

\def \bn{{\mathbfit{n}}}

\def \bB{\mathbfit{B}}

\def \bpi{\boldsymbol{\pi}}
\def \brho{\boldsymbol{\rho}}
\def \bxi{\boldsymbol{\xi}}

\def \bomega{\boldsymbol{\omega}}
\def \bnu{\boldsymbol{\nu}}

\newcommand{\w}{\boldsymbol}    % for vectors
\newcommand{\al}{\alpha}
\newcommand{\be}{\beta}
\newcommand{\de}{\delta}

\newcommand{\eps}{\epsilon}

\newcommand{\e}{\mathrm{e}}
\newcommand{\dd}{\partial} % частная производная
\newcommand{\ls}{\langle} % левая угловая скобка
\newcommand{\rs}{\rangle} % правая угловая скобка
\newcommand{\mean}[1]{\langle #1 \rangle}

\title[Suppression of small-scale dynamo]{Suppression of small-scale dynamo in time irreversible turbulence}

\author[A. V. Kopyev et al.]{
A. V. Kopyev,$^{1}$\thanks{E-mail: kopyev@lpi.ru}
A. S. Il'yn,$^{1,2}$\thanks{E-mail: asil72@mail.ru}
V. A. Sirota$^{2}$\thanks{E-mail: sirota@lpi.ru}
and K. P. Zybin$^{1,2}$\thanks{E-mail: zybin@lpi.ru}
\\
$^{1}$Lebedev Physical Institute of the Russian Academy of Sciences, Leninskij pr. 53, Moscow 119991, Russia\\
$^{2}$National Research University Higher School of Economics, Usacheva str. 6, Moscow 119048, Russia
}
\date{Accepted XXX. Received YYY; in original form ZZZ}

\pubyear{2023}

\begin{document}
\label{firstpage}
\pagerange{\pageref{firstpage}--\pageref{lastpage}}
\maketitle

\begin{abstract}
The conventional theory of small-scale magnetic field generation in a turbulent flow considers time-reversible random flows. However, real turbulent flows are known to be time irreversible: the presence of energy cascade is an intrinsic property of turbulence. We generalize the 'standard' model to account for the irreversibility. We show that even small time asymmetry leads to significant suppression of the dynamo effect at low magnetic Prandtl numbers, increases the generation threshold and may even make generation impossible for any magnetic Reynolds number. We calculate the magnetic energy growth rate as a function of the parameters of the flow.
\end{abstract}

\begin{keywords}
dynamo  --  turbulence -- Sun: magnetic fields -- methods: analytical
\end{keywords}

%%%%%%%%%%%%%%%%%%%%%%%%%%%%%%%%%%%%%%%%%%%%%%%%%%

%%%%%%%%%%%%%%%%% BODY OF PAPER %%%%%%%%%%%%%%%%%%

\section{Introduction}
\ \ The theory of magnetic field generation in a turbulent flow has
numerous applications and, in particular, is likely to explain the existence of magnetic field in
many astrophysical objects \citep{Parker, Brandenburg-rev, Sokoloff, Kuznetso, Kit}. The basic idea of turbulent dynamo
is that the magnetic lines stretch as they are carried by turbulent motion. However, a very wide
range of magnetized astrophysical objects of different scales implies very different parameters of
the medium.  A small-scale turbulent dynamo, i.e., a dynamo that takes place at scales much smaller
than the integral scale of turbulence \citep{Batchelor, FGV, Brandenburg12}, depends on  two basic parameters:
the hydrodynamic Reynolds number Re and the magnetic Reynolds number Rm. Their ratio Rm/Re=Pm is
called magnetic Prandtl number: it indicates whether the viscous scale $r_{\nu}$  of turbulence is
larger (Pm$>1$) or smaller than the resistive scale $r_d$. While both Rm and Re are large in cosmic
plasmas, their ratio varies crucially, and Pm is either very large or very small in astrophysical
objects. The Prandtl number controls the small-scale dynamo mechanism \citep{Batchelor, Rincon}.

The high-Pm turbulence is observed in interstellar and intergalactic media
\citep{Rincon, Han}. In these objects, the characteristic scale of magnetic field
generation is deep inside the viscous range of turbulence, so the stretching of magnetic lines is
exponential \citep{zeld}; the existence of the dynamo effect for this case is confirmed by different
theoretical approaches \citep{FGV, Chertkov99, Kazantsev} and by numerical simulations
\citep{schek-large-Pm, Brandenburg23}.

To the contrary, in stellar and planetary magnetism, in protostellar disks etc., one observes
low-Pm media Pm~$\sim 10^{-3}\div 10^{-7}$~\citep{Rincon, Rempel-rev}. In this case, small-scale dynamo can occur only in the
inertial range of turbulence, at scales $\rho \gtrsim r_d \gg r_{\nu}$. The existence of this
effect is still questionable. Direct numerical simulations (DNS) are very difficult  to perform in this
range of parameters \citep{Iskakov-Schek07, Brand-bottleneck}; some authors report the existence of generation, and
even the existence of a universal limit Rm$_c$ that guarantees the dynamo for any Re
\citep{Schek-Iskakov07, Warnacke}, although their estimates for the threshold Rm$_c$
differ essentially, 
and the values of Pm and especially Re that they achieve {(Pm$\gtrsim 10^{-3}$, Re$\lesssim 10^{5}$)} are still far from the observed in astrophysic objects.
Simulations performed by means of the shell model~\citep{Buchlin, Verma-Kumar, Frick-rev} and the implicit large-eddy simulations~\citep{Rempel-14, Hotta-16, Kapyla-18, Rempel-rev} 
allow to get small-scale dynamo at 
more realistic parameters, 
 however, 
the results 
 may depend significantly on the model
assumptions~\citep{Haugen, Petrosy-rev}. Observations 
of the Sun 
show the presence of small-scale magnetic field, but it is unclear
whether it is, indeed, generated in the inertial range by turbulent motion~\citep{Petrovay}, or it is merely a
result of fragmentation of the large-scale field produced by some other process; \citet{Stenflo-1, Stenflo-2} argues that the latter is more likely. Theoretical consideration  
predicts the
possibility of the dynamo effect in the low-Pm limit~\citep{Vain}.  
\citeauthor{Kazantsev} has found the dependence of the growth rate on the scaling
properties of isotropic incompressible  flow with infinite Re (infinite integral scale of
turbulence), under the assumption of Gaussian and delta correlated velocity statistics. In later
papers, the model was extended for the case of  finite Re~\citep{NRS, VK, Rog, Vinc, Arponen, Bold, Mal, Schober, IstKis}.

However, these theoretical works consider only time-reversible random flows. Actually, the 
time asymmetry is 
associated with the third-order velocity correlator, which is  zero in the classical Kazantsev
model. Meanwhile, the real turbulent flows are known to be time irreversible: 
this follows from the existence of
energy cascade~\citep{K41, Frisch}. The account of small time irreversibility in the
high-Pm limit has shown the decrease of 
all 
magnetic field 
statistical moments 
growth rate~\citep{ApJ, feedback, stationar}. So, there is a
reasonable question: how does it affect the magnetic field generation in the low-Pm case?

In this article, we consider the influence of the time irreversibility of
turbulence
%third order velocity correlator 
on the ability of
a low-Pm conductive fluid to generate magnetic field.
We find the instability criterion and show that\
the magnetic generation in a
turbulent
flow
may be suppressed completely if the third order correlator is large enough.  We show that in this case,
even for infinite Rm, there is no  generation for Pm  small enough but finite.

\section{Definitions and the $V^3$ model} 
The evolution of magnetic field in a (random) velocity field $\mathbfit{v}(\br,t)$ is described by the
equation
\begin{equation}\label{E:dynamicb}
\frac{\dd \mathbfit{B}(\mathbfit{r},t)}{\dd t} = \mathrm{rot} \bigl[\mathbfit{v}(\mathbfit{r},t)\times
\mathbfit{B}(\mathbfit{r},t)\bigr] + \eta \Delta \mathbfit{B}(\mathbfit{r},t)
\end{equation}
where $\eta$ is the magnetic diffusivity. The magnetic field energy in the problems under
consideration  is smaller than the kinetic energy of the flow at all scales, and the feedback
influence of the magnetic field on the velocity dynamics is proportional to $B^2$, so one can
neglect this feedback and consider $\bv (\br,t)$ as stationary random process with given
statistics.

We consider isotropic and homogenous stationary flow in incompressible fluid.  The Kazantsev
equation for the second-order magnetic
field correlator was derived  for 
Gaussian  $\delta$-correlated in time  velocity field,
\begin{equation} \label{E:vvt}
\mean{v_i(\mathbfit{r},t) v_j(\mathbfit{r}_1,t_1)} =D_{ij}(\mathbfit{r}-\mathbfit{r}_1)\,\de_\eps(t-t_1)
\end{equation}
Here $\de_\eps$ is not a 'physical' singularity but a regularized $\delta$ function: a narrow peak
with unitary square and width smaller than all physical scales. To avoid ambiguities, for any real
(short-correlated) velocity field one can define $D_{ij}$ by the integral
\begin{equation}
\label{E:D-def} D_{ij}(\w{\rho})=\int\mean{v_i(\mathbfit{r},t)
v_j(\mathbfit{r}+\w{\rho},t+\tau)}\mathrm{d} \tau
\end{equation}
(It does not depend on $\br$ and $t$ because of space and time homogeneity.) To take account of the
third-order velocity correlator, we introduce
\begin{equation} 
\label{E:X-def}
 F_{ijk}(\w{\rho}_1,\w{\rho}_2)=\int\mean{v_i(\mathbfit{r},t)
v_j(\mathbfit{r}+\w{\rho}_1,t_1)v_k(\mathbfit{r}+\w{\rho}_2,t_2)}\mathrm{d} t_1\mathrm{d} t_2
\end{equation} 
Now we use the $V^3$ model \citep{JoSS, ApJ, Spectrum}. It is a generalization of the Kazantsev-Kraichnan
model \citep{Kazantsev, Kraichnan-passive} and implies two assumptions: 1) the third order
correlator is assumed to be singular in time:
\begin{equation}   \label{E:vvvt}
\mean{v_i(\mathbfit{r},t) v_j(\mathbfit{r}+\w{\rho}_1,t_1) v_k(\mathbfit{r}+\w{\rho}_2,t_2)} =
F_{ijk}(\w{\rho}_1,\w{\rho}_2)
 \frac{\delta_{\eps}(t-t_1)\delta_{\eps}(t-t_2)+
\delta_{\eps}(t-t_1)\delta_{\eps}(t_1-t_2)+\delta_{\eps}(t-t_2)\delta_{\eps}(t_1-t_2)}{3} ,
\end{equation}
 and 2) the higher order correlators are set to zero.
We will discuss the validity of these assumptions a few strings later.

  Under these assumptions, we
derive the generalized Kazantsev equation (see Appendix~\ref{S:A} for derivation):
\begin{equation}\label{E:GenEq}
\begin{aligned}\frac\dd{\dd t}G=&\,2\eta\left( G''_{\rho\rho}+\frac{4G'_\rho}{\rho}\right)+b
G''_{\rho\rho}+\left(b'+4\frac {b}{\rho}\right)G'_{\rho}
+\left(b''+4\frac {b'}{\rho}\right)G+
\\
&
2cG'''_{\rho\rho\rho}+3\left(c'+\frac{4c}{\rho}\right)G''_{\rho\rho} +\left(3c''+\frac{12c'}{\rho} +\frac{8c}{\rho^2}+\frac{8d}{\rho}\right)G'_{\rho}+
\left(c'''+\frac{6c''}{\rho}+\frac{4c'}{\rho^2}-\frac{4c}{\rho^3}+\frac{4d'}{\rho}+\frac{12d}{\rho^2}
\right)G
\end{aligned}
\end{equation}
Here
\begin{equation}     
 G(\rho, t)=\langle  B_L(\mathbfit{r}+\w{\rho}, t) B_L(\mathbfit{r}, t) \rangle
\end{equation}
is the magnetic field correlator, and
\begin{equation}  \label{E:VCd-def}
\begin{array}{l}
b(\rho)= {n}_i  {n}_j  \left( D_{ij}(0)-D_{ij}(\w{\rho})\right),
\\  
c(\rho)=\tfrac12{n}_i  \Pi_{jk} F_{ijk}(\w{\rho},\w{\rho}),
\\  
d(\rho)= -\left(\tfrac18 n_i n_j \Pi_{kl} + \tfrac12n_j n_l \Pi_{ik}\right)   \frac{\dd}{\dd {\rho_1}_l}
F_{ijk}(\w{\rho}_1,\w{\rho})\bigl|_{\w{\rho}_1=\w{\rho}} ,
\\
\mathbfit{n}=\w{\rho}/\rho \ , \ \ \Pi_{ij}=\delta_{ij}-n_i n_j \ , \ \ B_L= \bB \cdot \bn
\end{array}
\end{equation}
represent the velocity correlators.

The applicability of the first assumption in long-term asymptotics for smooth velocity field
(Batchelor regime, $\rho < r_{\nu}$) was proved by~\citet{fin-time}. For rough velocity field
(inertial range, $\rho > r_{\nu}$) the correctness of this assumption remains a hypothesis: we
suppose that the effect of the finite correlation time is weaker than that of the asymmetry.
The insignificance of the finite-correlation time corrections for the Kazantsev model was confirmed
in DNS~\citep{Mason}. 

The second assumption produces non-physical artefacts,
as any non-Gaussian model with a finite number of cumulants does~\citep{Klyatskin, Mill, MY}.
Specifically, Eq. (\ref{E:GenEq}) contains the third order derivative, while the Kazantsev equation
($c=d=0$) is the second order differential equation. So, in addition to corrections to the two
'Kazantsev' modes, Eq. (\ref{E:GenEq})  provides a new eigenmode that is non-physical.  The account
of the whole set of the higher order correlators would destroy this 'parasite' mode, but one can
choose them small enough to have no significant effect on the two 'physical' modes.

For large Pm, it is enough to restrict the consideration by the viscous scales range.
Eq.(\ref{E:GenEq}) then reduces to the equation derived by~\citet{ApJ}. For small Pm, careful
analysis of the inertial range is needed.

\section{Rough velocity field} 
In the inertial range of hydrodynamical turbulence, $b(\rho)\propto \rho^{4/3}$; indeed, for
Kolmogorov turbulence~\citep{Vain, VK, Rog} one has
\begin{equation}
\delta v(\rho)\propto \rho^{1/3}, \,\,\tau_c(\rho)\propto \rho/\de v 
\propto\rho^{2/3}\, \Rightarrow \,\, b(\rho)\propto \de v^2 \tau_c\propto\rho^{4/3}
\end{equation}
Following \citeauthor{Kazantsev}, we hereafter consider a more general rough velocity field
with the power-law structure function $ b(\rho) \propto \rho^{1+s} $. From dimensional analysis we
then get 
\begin{equation}  \label{bcdpropto}
 \left\{\begin{aligned} &\delta v(\rho)\propto \rho^{s}
\\
&\tau_c(\rho)\propto\rho^{1-s}
\end{aligned}
\right.\Rightarrow\left\{\begin{aligned}
 b(\rho)&&\propto&& \rho^{1+s}
\\
c(\rho)&&\propto&& \zeta\, \rho^{2+s}
\\
d(\rho)&&\propto&& -  \beta\,\zeta\,\rho^{1+s},
\end{aligned}\right.
\end{equation}
Thus, we define
\begin{equation}\label{E:xi-Corrs}
\zeta = \frac{c(\rho)}{\rho \, b(\rho)} \ , \ \zeta \beta = - \frac{d(\rho)}{b(\rho)}
\end{equation}
The $\zeta$ and $\beta$ are dimensionless parameters that characterize the third order correlator
and that are constants inside the hydrodynamic inertial range of scales; $\zeta$ must be small to
ensure the applicability of the $V^3$ model, while $\beta$ is, generally, not restricted. From~\citet{JoT}
analysis of numerical simulations~\citet{JHTDB-1, JHTDB-3} it follows that~$\beta\simeq 2$.
We note that in (\ref{bcdpropto}),(\ref{E:xi-Corrs}) we neglect possible intermittency:
there is no evidence of anomalous scaling for time-integrated functions.

We also introduce the convenient length scale
\begin{equation}
r_d = \rho \left( 2 \eta / b(\rho)\right)^{1/(1+s)} \ , \rho \gg r_{\nu}
\end{equation}
For $\rho \gg r_d$, the molecular diffusion is negligibly small, and one can omit the first bracket
in (\ref{E:GenEq}). Then
\begin{equation}\label{E:EqXi}
\begin{aligned}
\frac 1{2\eta} \frac\dd{\dd t}G= &
\left(\frac{\rho}{r_d}\right)^{1+s} \left( G'' +\frac {5+s}{\rho}G'+\frac
{(4+s)(1+s)}{\rho^2}G\right)+
\\
&\zeta \, \left( \frac{\rho}{r_d}\right)^{1+s} 
\biggl( 2\rho\, G'''+({18+3s})G'' 
+\frac{3(5+s)(2+s)-8(\be-1)}{\rho}G'  \left.
+\frac{(4+s)\bigl((4+s)(1+s)-4\be\bigr)}{\rho^2}G\right)
 , \ \rho \gg r_d
\end{aligned}
\end{equation}

\section{Generation threshold} 
In the case $\zeta =0$, (\ref{E:GenEq}) turns into the Kazantsev equation.
The magnetic field generation in this model is possible for $s>0$, while for $s\le 0$ the magnetic
field decreases~\citep{Kazantsev}. Indeed, setting 
$\frac\dd{\dd t}G=0$  in~(\ref{E:EqXi}) we get two stationary modes, which for $\rho \gg r_d$ obey
the scaling law:
\begin{equation} \label{Kaz-sol}
G =  \rho^{-\frac{s+4}{2}} \left( C_+ \rho^{ \mu_{K+}} + C_{-} \rho^{ \mu_{K-}} \right)\ ,
\mu_{K\pm}= \pm\frac{\sqrt{3}}{2}\sqrt{-s(s+4)}
\end{equation}
 The exponential in time solutions
\begin{equation} \label{gamma-def}
G = e^{\Gamma t} G(\rho) \ ,  \dd G / \dd t = \Gamma G
\end{equation}
 of the Kazantsev equation exist if and only if the stationary solution oscillates in space.
(This can be derived from the oscillation theorem, keeping in mind that the Kazantsev equation
reduces to a Schr$\ddot{{\rm o}}$dinger equation; less formally, this implies that the solution
(\ref{Kaz-sol}) must match both with the 'viscous' solution at $\rho <r_d$ and with the large-scale
solution at $\rho \to \infty$.)    

We now use the same approach to find  the magnetic field generation threshold for
 $\zeta \ne 0$. Taking $\rho \gg r_d$
for the stationary mode of (\ref{E:EqXi})  we get the power-law solutions $ G \propto
\rho^{-\frac{s+4}{2}+ \mu_{\pm}}$ with the characteristic equation for the powers $\mu_{\pm}$:
\begin{equation} \label{chi}
\chi (\mu) =  \left( \mu^2 + \frac 34 s (s+4) \right) \left( 1+2\zeta \mu \right) - 8 \zeta
\beta \mu = 0
\end{equation}
Two of the solutions are close to those found for $\zeta=0$. The third solution is $\sim
\zeta^{-1}$, and it is a non-physical artefact. Again, the existence of generation corresponds to
oscillating stationary mode eigenfunctions, i.e., to imaginary roots of (\ref{chi}). The minimum of
the function $\chi (\mu)$, to the second order in $\zeta$, is 
\begin{equation}
\chi_{min} =  \frac 34 s (s+4) - \zeta^2 \left(  4\beta -  \frac 34 s (s+4)   \right)^2
\end{equation}
We see that the presence of $\zeta$ of any sign lowers the minimum. The generation is still
possible if $\chi_{min}>0$, i.e., if
\begin{equation} \label{zeta-crit}
|\zeta| < \zeta_{cr}=   \frac{ \sqrt{\frac 34 s (s+4)} }{
 |4\be -\frac 34 s (s+4)|}
\simeq  \frac{\sqrt{3s}}{4|\be|}
\end{equation}
For $s=1/3$, $\beta=2$ we get $\zeta_{cr}\simeq 1/7$. For larger $\zeta$, there is no dynamo.

We note that the role of $\beta$ is crucial: actually, one can see from (\ref{chi}) that the terms
$\zeta \beta$ are more effective to damp the oscillations than the terms with $\zeta$ alone. This
means that the contribution of $d(\rho)$ in (\ref{E:GenEq}) %,(\ref{E:d-def})
is more important than of $c(\rho)$.

\begin{figure}
\includegraphics[width=11cm]{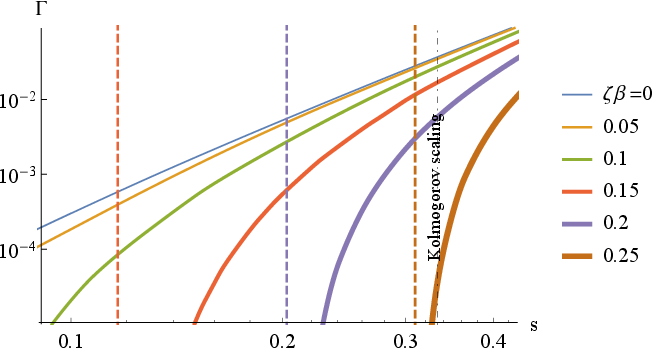}
\caption{The maximal growh rate $\Gamma$ as a function of $s$ for different values $\zeta\beta$ in
the minimal $V^3$ model; $\Gamma$ is normalized by $D Pm^{(1-s)/(1+s)}$ where $D$ is characteristic
frequency of small-scale eddies,  $D = - \frac 34 \left. \frac{\dd ^2}{\dd \rho^2} \langle
v_L(\brho) v_L({\bf 0})  \rangle \right| _{\rho=0}$.   } \label{fig1}
\end{figure}

\section{Growth rate }
 Even if the generation is not completely suppressed, its rate can be significantly slowed down.
To investigate the dependence of the generation rate on the time irreversibility, we focus on the
terms in~(\ref{E:GenEq}) that contain $d$, and set $c=0$. 
 We call this
 'minimal model'.
The estimate of $\zeta_{cr}$ obtained in the 'minimal model' coincides with (\ref{zeta-crit}) to
the first order in $s$.

We are looking for exponentially growing solutions~(\ref{gamma-def}). Now, the 'minimal model'
equation is a second-order ordinary differential equation containing the growth rate $\Gamma$;
it has a discrete spectrum of solutions for $\Gamma >0$ if $\zeta$ satisfies (\ref{zeta-crit}).

By means of the method proposed by~\citet[Appendix A]{blobs}, we find numerically  the maximal possible $\Gamma$ for given $s$ and
$\zeta$, assuming $\beta=2$.  We normalize it by $Pm^{\frac{1-s}{1+s}}$, so that it becomes
independent of the
regularization as Pm$\to 0$~\citep{Schober}.  
The results are presented in Fig.1. We see that they differ significantly from the Kazantsev case:
 for $s=1/3$ the generation suppression is essential even for $\zeta \beta \lesssim 0.2$.
 The vertical asymptotes in the logarithmic coordinates correspond to the generation threshold.
The dependence of the critical value of $\zeta \beta$ on $s$ is given in Fig.2. (For analytical
expression, see Appendix~\ref{S:B}.) It
 confirms a good agreement between the $V^3$ and the 'minimal' $V^3$ models (with and without
 $c(\rho)$) up to $s=1/3$. 
The whole analysis is performed in the limit of infinite integral scale, i.e.,  Rm$=\infty$; thus,
from Fig.2 it follows that, in the frame of our model, for $\zeta\beta \gtrsim 1/3$ generation is
impossible for any magnetic Reynolds number if Pm is small enough.

\section{Conclusion}
Summarizing, the presence of even small irreversibility is shown to suppress partially or
completely the small-scale magnetic field generation in the case of small Prandtl numbers. The mixed velocity and  gradients correlator $d(\rho)$ is
shown to play a crucial role in this anti-dynamo effect. 
The effect is strong enough to
be significant, e.g., for small-scale dynamo in convection zones of Sun and other stars, and perhaps in protostellar disks.
For large Pm (galaxies, galaxy clusters) the analogous effect is weaker and cannot suppress dynamo completely~\citep{ApJ}.

\begin{figure}[b]   
\includegraphics[width=11cm]{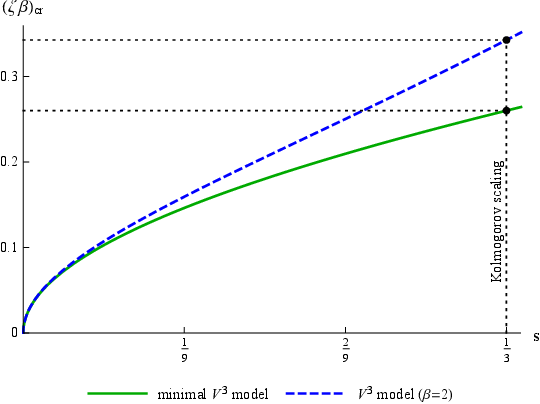}
\caption{\label{fig2} The critical value $(\zeta \beta)_{cr} (s)$ calculated for the 'minimal' $V^3$
model  and $V^3$ model (with $\beta=2$).}
\end{figure}

\section{Acknowledgments}
The authors are grateful to Professor A.V. Gurevich for his permanent attention to their work.
The authors are thankful to Professor L.L. Kitchatinov for drawing our attention to the problem of Solar magnetism.
We are also much obliged to A.M. Kiselev for providing the computation programs written by himself to collapse the tensor indexes.
This work of A.V. Kopyev was supported by the Foundation for the Advancement of Theoretical Physics and Mathematics (BASIS).

\section{Data availability}
The data underlying this article are available in the article.



\bibliographystyle{mnras}
\bibliography{example} % if your bibtex file is called example.bib

%\begin{thebibliography}{99}
%\end{thebibliography}

%%%%%%%%%%%%%%%%%%%%%%%%%%%%%%%%%%%%%%%%%%%%%%%%%%

%%%%%%%%%%%%%%%%% APPENDICES %%%%%%%%%%%%%%%%%%%%%

\appendix

\section{Generalized kazantsev equation}\label{S:A}

It is well-known that solenoidality and statistical homogeneity and isotropy in space restrict two point correlator to one scalar function: 
\begin{equation}
\ls B_i (\mathbfit{r}+\w{\rho}, t) B_j (\mathbfit{r}, t) \rs = G(\rho, t)\de_{ij}+ \frac\rho2\, G'_{\rho}(\rho, t) \,(\delta_{ij}-n_i n_j)
\end{equation}

Small-scale kinematical dynamo theory usually considers development of this correlator in time.
Kazantsev show that the closed equation on $G(\rho, t)$ can be derived when the velocity field is stationary, Gaussian and delta-correlated in time. In this case all statistical properties of the velocity field are governed by the pair correlator (\ref{E:D-def}):
\begin{equation}\label{S1}
D_{ij}(\mathbf{0})-D_{ij}(\boldsymbol{\rho})=b(\rho)\,\delta_{ij} + \frac\rho2\, b'(\rho) \,(\delta_{ij}-n_i n_j)
%\eqno{(S1)}
\end{equation}

To derive a dynamical equation on $G(\rho, t)$ in the frame of '$V^3$ model' two additional two-point correlators are needed:
\begin{align}%\label{E:F-def}
&F_{ijk}(\w{\rho},\w{\rho})=\iint\ls v_i(\mathbfit{r},t) v_j(\mathbfit{r}+\w{\rho},t+\tau_1)v_k(\mathbfit{r}+\w{\rho},t+\tau_2)\rs\mathrm{d} \tau_1\mathrm{d} \tau_2,
\\%\label{E:Y-def}
&\frac{\dd}{\dd {\rho_1}_l} F_{ijk}(\w{\rho}_1,\w{\rho})\bigl|_{\w{\rho}_1=\w{\rho}}=\iint\ls v_i(\mathbfit{r},t) A_{jl}(\mathbfit{r}+\w{\rho},t+\tau_1)v_{k}(\mathbfit{r}+\w{\rho},t+\tau_{2})\rs\mathrm{d} \tau_1\mathrm{d} \tau_2
\end{align}
where $A_{jl}=\dd v_j/{\dd r_l}$ is a velocity gradient tensor. The first of the correlators involves one new
scalar function $c(\rho)$~\citep{LL}:
\begin{equation}\label{S2}
F_{ijk}(\boldsymbol{\rho},\boldsymbol{\rho})=c(\rho)n_i \de_{jk}-\left(c(\rho)+\frac\rho2 \, c'(\rho)\right)\left(n_j \de_{ik}+n_k\de_{ij}\right)-\left(c(\rho)- \rho\, c'(\rho)\right)n_in_jn_k
\end{equation}

The second one requires one more scalar function $d(\rho)$~\citep{JoT}:

\begin{align}\label{S3}
\frac{\dd}{\dd {\rho_1}_l} F_{ijk}(\w{\rho}_1,\w{\rho})\bigl|_{\w{\rho}_1=\w{\rho}}=&-\frac{c(\rho)}{2 \rho} \de_{i l} \de_{j k} - 
 d(\rho) n_{i} n_{k} \de_{j l} + 
 \left( d(\rho) - \frac{c'(\rho)}2 
    \right)\de_{i k} \de_{j l}  + 
\left(\frac{c(\rho)}{\rho}  - d(\rho) + 
     c'(\rho)\right) \de_{i j} \de_{k l}+
\\
&\notag
+ \left(\frac{c(\rho)}{\rho}  +  d(\rho) -  c'(\rho)\right) n_{i} n_{j} \de_{k l}-  \left(\frac{c(\rho)}{2\rho}  + 2 d(\rho) - \frac{c'(\rho)}2 \right)  n_{j}  n_{l} \de_{i k} +
\\
&\notag
+ \left(\frac{c(\rho)}{2\rho}  - \frac{c'(\rho)}{2}\right)\left(n_{i} n_{l} \de_{j k} + n_{j} 
   n_{k} \de_{i l} \right) -  \left(\frac{3 c(\rho)}{2\rho}  - \frac{3 c'(\rho)}{2} + 
    \frac{\rho c''(\rho)}{2}\right) n_{i} n_{j} n_{k}  n_{l}+
\\\notag
&
 +\left(-\frac{c(\rho)}{2\rho}  + 2 d(\rho) + \frac{c'(\rho)}{2} + 
    \frac{\rho c''(\rho)}{2}\right)n_{k} 
   n_{l} \de_{i j} 
\end{align}%\eqno{(S3)}

Let us introduce the notation:
\begin{align}
B_\al = B_\al(\mathbfit{r},t), \quad B'_\al = B_\al(\mathbfit{r}',t), \quad
v_\al=v_\al(\mathbfit{r},t)&, \quad v_\al^1 = v_\al(\mathbfit{r}_1,t), \quad
v_\al^2 = v_\al(\mathbfit{r}_2,t), \quad
\\
\dd_\al = \frac{\dd}{\dd r_\al}, \quad\dd'_\al = \frac{\dd}{\dd r'_\al}, \quad\dd^{\rho}_\al = \frac{\dd}{\dd \rho_\al}&, \quad\dd^1_\al = \frac{\dd}{\dd {r_1}_\al}, \quad\dd^2_\al = \frac{\dd}{\dd {r_2}_\al}
\end{align}

Then from the induction equation (1) one can obtain:
\begin{equation}
\frac{\dd}{\dd t} \ls B_\al B'_\be \rs  = 
-\e_{\al mn} \e_{n p q} \dd_m^{\rho} \ls v_p B_q B'_\be \rs - \e_{\be mn} \e_{n p q} \dd_m^{\rho} \ls v_p B_q B'_\al \rs 
+ 2\eta\, \dd^{\rho}_m \dd^{\rho}_m \ls B_\al B'_\be \rs
\end{equation}
where $\e_{n p q}$ is Levi-Civita symbol.
The latter equation is non-closed. However the procedure of splitting~\citep{Klyatskin} can make it the closed one. A detailed performance of this procedure in the frame of $V^3$ model is given by~\citet[Appendix B]{ApJ} for smooth velocity field. The case of rough velocity field is quite analogous, so we just point out some key moments and differences between two cases. The splitting procedure is based on Furutsu-Novikov formula, which can be significantly simplified in the frame of $V^3$ model  ($\de/\de v_i$ means a functional derivative):
\begin{align}
\ls v_p B_q B'_r\rs & = \sum_{N=1}^\infty \frac{1}{N!}
\int \, d\mathbfit{r}_1 dt_1 \dots d\mathbfit{r}_N dt_N \ls v_p(\mathbfit{r},t) v_{i_1}(\mathbfit{r}_1,t_1) \dots v_{i_N}(\mathbfit{r}_N,t_N)\rs_c\ls \frac{\de^N \bigl(B_q(\mathbfit{r},t) B_r(\mathbfit{r}',t)\bigr)}{\de v_{i_1}(\mathbfit{r}_1,t_1) \dots \de v_{i_N}(\mathbfit{r}_N,t_N)} \rs
\\\notag
&=\int \, d\mathbfit{r}_1 dt_1 \ls v_p(\mathbfit{r},t) v_{i_1}(\mathbfit{r}_1,t_1)\rs\,\ls \frac{\de \bigl(B_q(\mathbfit{r},t) B_r(\mathbfit{r}',t)\bigr)}{\de v_{i_1}(\mathbfit{r}_1,t_1)}\rs
\\\notag
&\,\,\,\,\,\,\,\,\,\,\,\,\,\,\,\,\,\,\,\,+\frac{1}{2}\int \, d\mathbfit{r}_1 dt_1d\mathbfit{r}_2 dt_2 \ls v_p(\mathbfit{r},t) v_{i_1}(\mathbfit{r}_1,t_1) v_{i_2}(\mathbfit{r}_2,t_2)\rs\,\ls \frac{\de^2 \bigl(B_q(\mathbfit{r},t) B_r(\mathbfit{r}',t)\bigr)}{\de v_{i_1}(\mathbfit{r}_1,t_1)\de v_{i_2}(\mathbfit{r}_2,t_2)}\rs
\\\notag
& = 
\frac12 \int \, d\mathbfit{r}_1 D_{pj}(\mathbfit{r}-\mathbfit{r}_1) \left(\ls\tfrac{\de B_q}{\de v_j^1} B'_r \rs+\ls B_q\tfrac{\de  B'_r}{\de v_j^1} \rs\right)
\\\notag
&\,\,\,\,\,\,\,\,\,\,\,\,\,\,\,\,\,\,\,\,+\frac16 \int \, d\mathbfit{r}_1 d\mathbfit{r}_2 X_{pjk}(\mathbfit{r}-\mathbfit{r}_1, \mathbfit{r}-\mathbfit{r}_2) \left(\ls\tfrac{\de^2 B_q}{\de v_j^1 \de v_k^2 }B'_r\rs+\ls\tfrac{\de B_q}{\de v_j^1 }\tfrac{\de B'_r}{\de v_k^2 }\rs+\ls\tfrac{\de B_q}{\de v_k^2 }\tfrac{\de B'_r}{\de v_j^1 }\rs+\ls B_q\tfrac{\de^2 B'_r}{\de v_j^1 \de v_k^2 }\rs\right)
\end{align} 

Finding the functional derivatives (see~\cite{ApJ}) one obtains the following equation: 
\begin{align}
\frac{\dd}{\dd t} \ls B_\al B'_\be \rs  = 2\eta\, \dd^{\rho}_m \dd^{\rho}_m \ls B_\al B'_\be \rs-\,\,\,\,\,\,\,\,\,\,\,\,\,\,\,\,\,&
\\
-\frac{1}{2}\bigl(\e_{\al mn} \e_{n p q} \de_{\be r}+ \e_{\be mn} \e_{n p q} \de_{\al r}\bigr)\e_{j_1 j k_1}\dd^{\rho}_m\dd^{\rho}_{i_1}  &\left(-\e_{q i_1 j_1} D_{pj}(\mathbf{0}) \ls B_{k_1} B'_r\rs +\e_{r i_1 j_1}  D_{pj}(\boldsymbol{\rho}) \ls B_q B'_{k_1}\rs\right)-
 \notag
 \\
-\frac{1}{6}\bigl(\e_{\al mn} \e_{n p q} \de_{\be r}+ \e_{\be mn} \e_{n p q} \de_{\al r}\bigr) \e_{j_1 j k_1}\e_{j_2 k k_2}&\times
\notag
\\
\times\dd^{\rho}_m\Bigl(2\e_{q i_1 j_1} \e_{r i_2 j_2}
&\dd^{\rho}_{i_2}\bigl(\tfrac{\dd}{\dd {\rho_1}_{i_1}} F_{kjp}(\w{\rho}_1,\w{\rho})\bigl|_{\w{\rho}_1=\w{\rho}} \ls  B_{k_1}  B'_{k_2}\rs-F_{kjp}(\boldsymbol{\rho})  \dd^{\rho}_{i_1}\ls  B_{k_1}  B'_{k_2}\rs\bigr)+
 \notag
\\
+\e_{r i_1 j_1}\e_{k_1 i_2 j_2} 
&\dd^{\rho}_{i_1} \bigl(\tfrac{\dd}{\dd {\rho_1}_{i_2}} F_{pkj}(\w{\rho}_1,\w{\rho})\bigl|_{\w{\rho}_1=\w{\rho}} \ls B_{q}B'_{k_2}\rs-F_{pkj}(\boldsymbol{\rho}) \dd^{\rho}_{i_2}\ls B_{q}B'_{k_2}\rs \bigr)\Bigr)
\notag 
\end{align}

Substituting (\ref{S1}), (\ref{S2}) and (\ref{S3}) into the latter expression and convoluting enormous number of summands with the aid of computer
algebra, one arrive at the generalized Kazantsev equation (\ref{E:GenEq}). 

Note that in the case of smooth velocity field considered by~\citet{ApJ} functions $c$ and $d$ are bounded by the symmetries of the flow: $c(\rho)/d(\rho)=-\rho/6$ or $\be=6$. In the case of the rough field kinematical reasons are insufficient to obtain the relation between two functions~\citep{JoT}. If one suppose that the correlation times of $\ls v_i(\mathbfit{r},t) v_j(\mathbfit{r}+\w{\rho},t+\tau_1)v_k(\mathbfit{r}+\w{\rho},t+\tau_2)\rs$ and $\ls v_i(\mathbfit{r},t) A_{jl}(\mathbfit{r}+\w{\rho},t+\tau_1)v_k(\mathbfit{r}+\w{\rho},t+\tau_2)\rs$ coincide, the numerical simulation data gives $\be\simeq2$~\citep{JoT}.

\section{Critical antidynamo irreversibility in V3 model}\label{S:B}

It is convenient to introduce a notation
\begin{equation}
\sigma=\frac{3s(s+4)}{4}
\end{equation}

Consider~(\ref{chi}). The minimum of cubic parabola must have a non-negative value to have no generation. The following condition $\zeta>\zeta_{cr}(s,\be)$ can be obtained analytically. 
In minimal $V^3$ model ($\zeta\be\gg\zeta$) the cubic parabola degenerates into the quadratic one and the solution is obvious:
\begin{equation}\label{B2}
(\zeta\be)_{cr}=\frac{\sqrt{\sigma}}{4}
\end{equation}

In general $V^3$ model the analitycal solution can also  be found:
\begin{equation}\label{B3}
\zeta_{cr}=\frac{\sqrt{\sigma^2+10\be\sigma-2\be^2+2\sqrt{\be}(\be+2\sigma)^{3/2}}}{2(4\be-\sigma)^{3/2}}
\end{equation}

Dependences (\ref{B2}), (\ref{B3}) are depicted in Figure~\ref{fig2}.

%%%%%%%%%%%%%%%%%%%%%%%%%%%%%%%%%%%%%%%%%%%%%%%%%%

% Don't change these lines
\bsp	% typesetting comment
\label{lastpage}
\end{document}